\documentclass[fleqn,10pt]{wlscirep}
\usepackage[utf8]{inputenc}
\usepackage[T1]{fontenc}
\title{Greedy control of cascading failures in interdependent networks}

\author[1,*]{Malgorzata Turalska}
\author[1]{Ananthram Swami}
\affil[1]{CCDC Army Research Laboratory, Network Science Division, Adelphi, MD 20783, USA}
\affil[*]{mg.turalska@gmail.com}

\begin{abstract}
Complex systems are challenging to control because the system responds to the controller in a nonlinear fashion, often incorporating feedback mechanisms. Interdependence of systems poses additional difficulties, as cross-system connections enable malicious activity to spread between layers, increasing systemic risk. In this paper we explore the conditions for an optimal control of cascading failures in a system of interdependent networks. Specifically, we study the Bak-Tang-Wiesenfeld  sandpile model incorporating a control mechanism, which affects the frequency of cascades occurring in individual layers. This modification allows us to explore sandpile-like dynamics near the critical state, with supercritical region corresponding to infrequent large cascades and subcritical zone being characterized by frequent small avalanches. Topological coupling between networks introduces dependence of control settings adopted in respective layers, causing the control strategy of a given layer to be influenced by choices made in other connected networks. We find that the optimal control strategy for a layer operating in a supercritical regime is to be coupled to a layer operating in a subcritical zone, since such condition corresponds to reduced probability of inflicted avalanches. However this condition describes a parasitic relation, in which only one layer benefits. Second optimal configuration is a mutualistic one, where both layers adopt the same control strategy. This work demonstrates that control protocols in systems of interdependent networks need to take into account higher-order organization of the system and cannot be designed independently, maximizing benefits only for their individual layers. 
\end{abstract}
\begin{document}

\flushbottom
\maketitle

\thispagestyle{empty}

\section*{Introduction}

Networks constitute the theoretical framework behind structural and dynamical properties of a plethora of natural and man-made systems \cite{Newman2010,barabasi2016network}. Evolution of social, communication and information platforms, spread of infectious diseases or dependencies between financial markets are all captured by network science concepts. In particular, the above examples are all demonstrations of a property characteristic to networks, namely their ability to propagate perturbations. Cascading failures form an especially important class of such perturbations, as they capture processes of malfunction spreading. Starting from a localized failure, interactions between components of a network initiate a domino-like effect, resulting in catastrophic events such as blackouts in power grids \cite{Carreras2002,Weng2006,Hines2009}, crashes in financial markets \cite{Battiston2012,Bardoscia2017} and  extinctions in ecological systems \cite{Estes2011,Motter2011}. The massive failures are only amplified by the interdependencies between networks \cite{Havlin2010}, making inquiries into control protocols  reducing the risk of cascading failures an active area of research.     

In particular, numerous investigations into topological features of interconnected networks have shown that structural connectivity plays significant role in reducing vulnerabilities of those systems \cite{Havlin2010}. Next to traditional features such as degree distribution, higher-order network organization, reflected by intra- and inter-layer degree correlations, has been identified as a necessary condition for structural stability and robustness of coupled networks \cite{Reis2014, Kleineberg2017}. Similar observations have been made regarding dynamical processes on networks. Recent studies have determined that the density of interlayer connections affects occurrence of cascading failures in interconnected systems \cite{Brummitt2012,turalska2019}, while correlations present in the multilayer structure facilitate cooperation in evolutionary games \cite{Kleineberg2017} or decrease of epidemic thresholds \cite{mendiola2012}. Those studies however focus on the impact the structure of a system has on observed dynamics, while in many realistic scenarios large scale changes to the network structure are not a preferable or available method of exerting control. More practical approaches focus on early detection and prediction of cascading events, and ultimately on control of events after they have been triggered \cite{Motter2017b}. In this paper we develop a systematic framework addressing the latter case. 

We model cascading failure process with the Bak-Tang-Weisenfeld (BTW) sandpile model of self-organized criticality \cite{Bak1987,Bak1988}. The BTW process is an archetypal model for the cascades of load, in which the distribution of event sizes is characterized by a power law. Such scaling of failure cascades is observed e.g. in electrical blackouts \cite{Carreras2002, Weng2006}, earthquakes \cite{Geller1997,Sornette2004} and forest fires \cite{sinha2000}, making the BTW model a valuable tool for studying propagation of failures resulting from a malfunction of a single element in a system. Within the framework of the BTW model, control can be implemented as any action interfering with the process of load accumulation on a network, resulting in reduction of the probability of extreme events. The principle behind such action follows the reasoning behind e.g. triggering snow avalanches in order to prevent snow accumulation that could result in deadly avalanches, or igniting controlled forest fires in order to reduce the probability of uncontrolled events spreading over wide areas. 

The first proposed control scheme of the BTW process is based on triggering cascades on sites that are near to becoming seeds of cascading events \cite{cajueiro2010,Cajueiro2010b}. This approach leads to a redistribution of load on the network through small scale cascades, and it prevents the occurrence of massive failures. More recently No\"{e}l \textit{et. al.} \cite{Noel2013} defined a broader scheme by controlling how often cascades of any size occur in the system. Depending on the control drive this protocol recovers earlier results \cite{cajueiro2010,Cajueiro2010b} by avoiding large cascades at the cost of causing many small ones. However it also allows for an opposite situation, where cascades of any size are avoided at all cost. The latter rule leads to accumulation of load on the network, which in turn results in very infrequent, but highly destructive failures. 

As mentioned earlier, pervasiveness of interconnected systems determines the need to extend control schemes proposed for individual networks to more complex structures. Thus in this paper we discuss control of cascading failures extending the protocol of No\"{e}l \textit{et. al.} to a system of interdependent networks. Our goal is to identify conditions for optimal control exerted through modification of the dynamical rules defining cascading process. We observe that the topological coupling between networks introduces dependence of control settings adopted in respective layers, causing the control strategy of a given layer to be influenced by choices made in other connected layers. Therefore due to cascading events that propagate across layers, a strategy optimal for an individual network becomes suboptimal in the connected system. Thus we demonstrate that  control protocols in systems of interdependent networks need to take into account higher-order organization of the system and cannot be designed independently, maximizing benefits only for the individual layers.  

\section*{Methods and models}

\subsection*{Sandpile model} 
The BTW model is an idealized model of cascading dynamics driven by load redistribution on a network \cite{Pruessner2012}. In this paper we consider a generalization of the classical BTW dynamics \cite{Bak1987,Bak1988} into a network of arbitrary topology, as follows \cite{Brummitt2012,Noel2013}. We consider a network of $N$ nodes, where each node has a fixed capacity to hold grains of sand. The capacity of a node is the maximal amount of sand that it can hold and it is set to $k-1$, where $k$ is the degree of a node. Thus, a $(k-1)$-sand node of degree $k$ is said to be \textit{at capacity}, and adding sand to such node brings it \textit{over capacity}, initiating a toppling event. 

The dynamics of sandpile model consist of a slow and a fast process. The slow process adds one grain of sand to a node chosen uniformly at random. If this addition brings the initial node over capacity, the fast process of the propagation of an avalanche is initiated. The initial node topples and sheds one gain of sand to each of its neighbors. This shedding might result in neighboring nodes becoming overloaded. Thus any node that exceeds its capacity topples, in the same way as the first node, distributing its load to its neighbors, who then may topple as well. This repeated shedding can result in a cascade of events, which continues until all nodes are below or at their capacity. Once equilibrium is restored, the slow process proceeds with adding another grain of sand. In order to prevent the system from becoming saturated with sand, we incorporate a dissipation mechanism into the dynamics. Whenever a grain of sand is moved between nodes during a shedding event, it dissipates (is removed) with a small probability $f$. In all numerical simulations presented in this paper $f=0.05$.  
The \textit{size} $s$ of a cascade is defined as the total number of toppling events initiated by deposition of a single grain of sand. The mean-field solution to the BTW model is characterized by a distribution of cascade sizes, $P(s)$, that exhibits a power law scaling with exponent $-3/2$ \cite{Pruessner2012}. The same scaling describes the size of cascading events observed for a wide range of networks, from random regular networks \cite{Brummitt2012}, through networks with narrow degree distributions \cite{Bonabeau1995}, to particular classes of scale-free topologies \cite{Goh2003}.  

\begin{figure}[ht!]
\centering
\includegraphics[width=\linewidth]{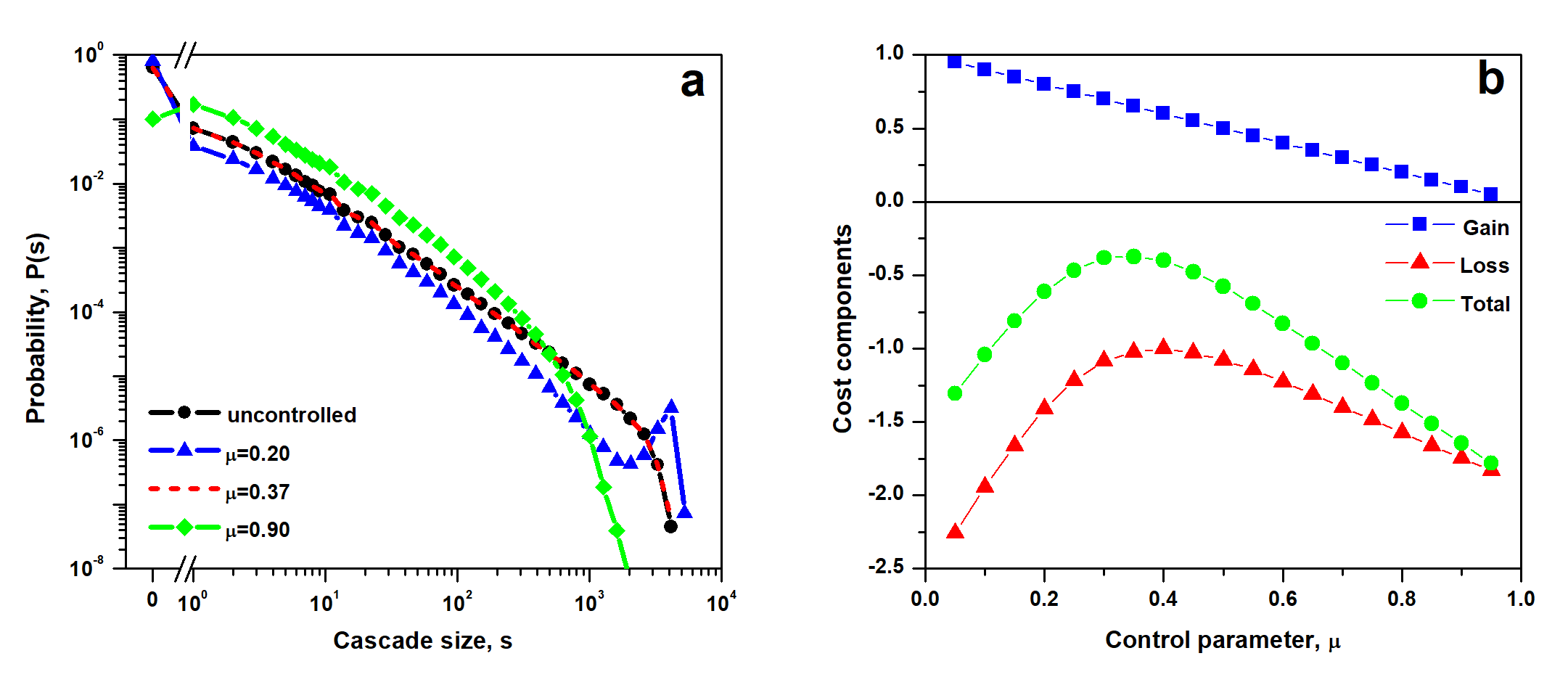}
\caption{ Control parameter $\mu$ shapes the cascade size distribution, $P(s)$, and as a result it determines the value of average cost, $\langle C \rangle$, associated with given control strategy. (a) For a single network the chance of a non-zero size cascade is $P(s>0)=\mu$ and the complementary probability of no cascade is $P(s=0)=1-\mu$. The native BTW model corresponds to the sandpile dynamics simulated with control parameter $\mu^\ast=0.37$ (black circles and red dashed line). Values of $\mu<\mu^\ast$ lead to supercritical dynamics characterized by increased probability of cascades equal to the system size (blue triangles), while values of $\mu>\mu^\ast$ result in a subcritical state, where probability of smaller size failures is increased (green diamonds). (b) System gains are proportional to the frequency at which cascades are avoided, thus $G(\mu)=1-\mu$ (blue squares). Cost associated with non-zero size cascades grows nonlinearly with cascade size, $C(s)\sim s^\alpha$ (red triangles), and parameters are selected such that average total cost is maximized at $\mu=\mu^\ast$ (green dots). Figures denote results of numerical simulations on random regular $R(4)$ graph with $N=5000$, $\langle k \rangle=\delta(k-4)$ and dissipation parameter $f=0.05$. The cost function is $C(s)=\frac{1}{2}s^{3/4}$. }
\label{fig:single_net}
\end{figure}

\subsection*{System topology} 
In this paper we consider the BTW process on a random $4-$regular graph, R(4), (a random network of degree-four nodes). As mentioned earlier, despite its simplicity, such topology shares the functional form of the statistics of cascade size, $P(s)$, with numerous other graphs. Additionally random regular graphs well approximate the real topologies of networks such as power grids \cite{Brummitt2012}, making them a useful basic structure to explore. 

To study the sandpile dynamics on interconnected networks, we generate independently two R(4) networks of the same size and connect nodes between them in a random fashion. We then refer to those original networks as layers A and layer B, respectively. The parameter $p$, which varies between $0$ and $0.50$, dictates the density of coupling between the layers. The value of $p$ affects the degree distribution characterizing individual layers, which evolves from $P(k)=\delta(k-4)$ at $p=0$ to $P(k)=(1-p)\delta(k-4)+p\delta(k-5)$ at $p>0$. Thus at any $p>0$ individual layers are no longer pure random $4-$regular graphs. Following the sandpile dynamics outlined earlier, the capacity of nodes to hold sand is set to $(k-1)$, and thus $5-$degree nodes have larger capacity than $4-$degree nodes.

\subsection*{Control mechanism}
Since in numerous realistic scenarios exerting control through adjustments or changes to the network structure is not possible, in this paper we consider a protocol affecting solely the dynamics of the sandpile process. However we keep the main features of the process unchanged. Namely the cascading mechanism of redistribution of sand from overloaded nodes and the dissipation rate are unaffected by the control protocol. The only remaining degree of freedom is the deposition of an initial grain of sand, which in the original sandpile model is done randomly. Thus rather than depositing a grain of sand on a randomly selected node, what might or might not result in a cascade, one sets to control the probability of initializing a cascading event. This is realized by an informed selection of the node to which a grain of sand is being added. With probability $\mu$ we deposit sand on a node that is at capacity (what causes a cascade) and with probability $1-\mu$ we select a node that is below capacity (no cascade). 

As demonstrated by No\"{e}l \textit{et. al.} \cite{Noel2013}, the native sandpile model is recovered by setting $\mu=\mu^\ast$, where particular value of $\mu^\ast$ depends on system size and adopted dissipation. Furthermore, as shown on Fig. \ref{fig:single_net}a, departure from the condition of $\mu=\mu^\ast$ allows us to explore regions near the self-organized critical state. The setting of $\mu > \mu^\ast$ corresponds to preferentially depositing sand on nodes that are at capacity, thus initiating cascades more frequently then in the native case. This strategy results in numerous small cascades and decreased occurrence of large size events, what is reminiscent of a subcritical state. An opposite strategy, $\mu < \mu^\ast$, amounts to avoiding cascades at all cost by allocating the first grain of sand to nodes that are below their capacity. This choice however leads to accumulation of the sand on the network, since the dissipation mechanism acts only during evolution of a cascading process. As a result despite the fact that the average rate of failures drops, the system occasionally suffers from rare but extremely large events. Thus the system is in a supercritical-like state, with maximum cascade size reaching the size of the network. 

\begin{figure}[!ht]
\centering
\includegraphics[width=\linewidth]{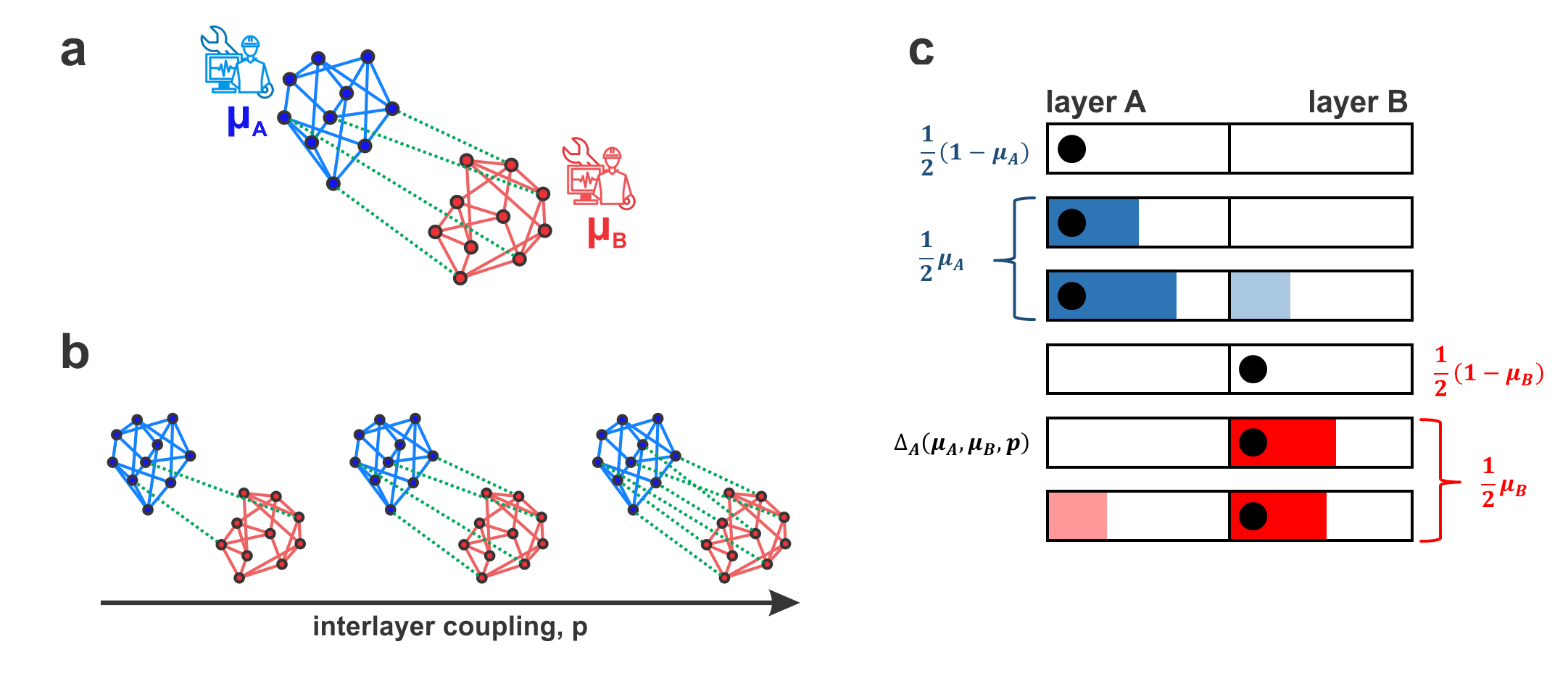}
\caption{ (a) Interdependency of systems poses challenges to the controllers operating in the individual networks. (b) Increasing the number of interlayer connections increases the chance for cascading failures to  propagate from one layer to the other one. (c) In a system of coupled networks the probability of no cascade occurring in layer $A$ is $\frac{1}{2}(1-\mu_A)+\frac{1}{2}(1-\mu_B)+\frac{1}{2}\mu_B\Delta_A$, demonstrating the coupling between controller settings chosen by respective layers. Schematic figure illustrates how frequency of particular events (no cascade, cascade originating in layer A and propagating or not to layer B, etc.) is tied to the control parameter chosen in each layer. Black dot denotes a grain of sand deposited in a given layer, colored areas denote non-zero size cascade. } 
\label{fig:schematic}
\end{figure}

In the system composed of two interconnected networks, the settings of the controller are chosen independently for each layer. At each instance of the slow part of the sandpile dynamics the layer on which a grain of sand is deposited is chosen at random. Next, if layer A is picked, with probability $\mu_A$ the grain is deposited on \textit{at-capacity} node belonging to that layer. If layer B is selected, deposition of sand on \textit{at-capacity} node of layer B happens with probability $\mu_B$. This approach allows us to specify the occurrence of cascades in individual layers and to observe how that choice is affected by coupling between layers.

\subsection*{Measure of risk}
the notion of control is naturally associated with perception of risk, and thus a measure of cost is associated with events occurring in the system. Since numerous complex systems, such as infrastructure or transportation networks, are characterized  by nonlinear and non-local interactions, it's natural to assume that the cost of cascading events is a nonlinear function of event size. Thus we define a  cost function $C(s)$ as a generic nonlinear function of the cascade size
\begin{equation}
  C(s) =
    \begin{cases}
      +1           & \text{if s=0 }\\
      -cs^{\alpha} & \text{if s>0 },
    \end{cases}    
\end{equation}
capturing the fact that large cascades inflict disproportionately greater damage than small events, when immediate and intermediate changes to system function are taken into consideration. The negative sign of the cost function associated with non-zero size cascades reflects the detrimental effect failures have on system functionality. Avoidance of cascading failures is considered beneficial to the system and thus incurs a positive cost of $+1$ for zero size cascades. Since the probability of zero size cascades is directly defined by the control parameter, $P(s=0)=1-\mu$, the average cost associated with given value of $\mu$ is
\begin{equation}
    \langle C(\mu) \rangle=\sum_{s=0}^{\infty} P(s)C(s)=P(s=0)C(s=0)+ \sum_{s>0}^{\infty} P(s)C(s)=1-\mu + \sum_{s>0}^{\infty} P(s)C(s).
\end{equation}
the first component denotes the gains due to avoidance of cascades, while the second term defines losses due to propagating failures. The balance between both terms determines to what extent adopted control setting is beneficial or detrimental to the system. 

\subsection*{Risk in interconnected networks}

Interdependency of networks determines systemic risk, as actions of controllers operating in the individual layers reverberate across the system through interlayer connections. In the case of sandpile dynamics discussed in this paper, the control strategy determined by the choice of $\mu$ value in one layer will be affected by control settings in the other layer. As illustrated on Figure \ref{fig:schematic}a, we consider two controllers, each selecting a control protocol for its layer, independently of decisions made in the other layer. However with growing interlayer connectivity (Fig. \ref{fig:schematic}b) cascades of failures might more likely spread across connected networks, modifying the original control settings and increasing risk beyond the value set by the controller operating the network.   

We start from a configuration of two random regular networks, one operating with control setting $\mu_A$, and the second working with control setting $\mu_B$. The schematic presented in Figure \ref{fig:schematic}c outlines possible scenarios occurring during the sandpile process dynamics. The black dot denotes allocation of the initial grain of sand and color blocks illustrate cascades that might be a result of this deposition. The plot distinguishes between two layers, A and B, and allows for tracking how the interlayer coupling is influencing control settings set for each layer. In both layers, deposition of the grain of sand does not lead to a cascade, with probability $\frac{1}{2}(1-\mu_A)$ and $\frac{1}{2}(1-\mu_B)$, respectively for layer A and B. Naturally then, a cascade occurs with probability $\frac{1}{2}\mu_A$ and $\frac{1}{2}\mu_B$ in each layer. However this event consists of two separate cases: one in which a cascade originates in layer A (B) and is contained in that layer, and second one in which a cascade is initiated in layer A (B) and spreads to layer B (A) through interlayer connections. The rate of those events is $\Delta_B$ ($\Delta_A$) and $\theta_B$ ($\theta_A$), respectively, where $\theta_A=1-\Delta_A$ and $\theta_B=1-\Delta_B$. In general those rates are a function of both control settings $\mu_A$ and $\mu_B$ and interlayer coupling $p$.

Now let's focus on events occurring in layer A, as corresponding relations can be easily written for layer B. We do not observe cascades in layer A in three cases: when a grain of sand is dropped in layer A and does not causes a cascade, when a grain of sand is dropped in layer B and does not lead to a failure and when a grain of sand dropped in layer B causes a cascade that is entirely contained in that layer. Thus the probability of observing cascade of size zero in layer A is:
\begin{equation}
    \label{s_zero}
    P(s_A=0)=\frac{1}{2}(1-\mu_A)+\frac{1}{2}(1-\mu_B)+\frac{1}{2}\mu_B\Delta_A(\mu_A,\mu_B,p).
\end{equation}
Consecutively, non-zero events occur with probability:
\begin{equation}
    \label{s_non_zero}
    P(s_A>0)=\frac{1}{2}\mu_A+\frac{1}{2}\mu_B\theta_A(\mu_A,\mu_B,p)=\frac{1}{2}\mu_A+\frac{1}{2}\mu_B\Big(1-\Delta_A(\mu_A,\mu_B,p)\Big).
\end{equation}
The influence that the control setting chosen by layer B  exerts on the sandpile dynamics observed in layer A is clearly visible through these expressions. Since the expected total cost incurred by layer A is 
\begin{equation}
    \label{average_cost}
    \langle C_A \rangle=\sum_{s_A=0}^{\infty} P(s_A)C(s_A)=P(s_A=0)C(s_A=0)+ \sum_{s_A>0}^{\infty} P(s_A)C(s_A),
\end{equation}
both gains and losses are a function of the $\mu$ values selected by layer A and B. 

In the following section, we will discuss in detail how particular choices of control parameters and strength of interlayer coupling affect the risk perceived by individual layers and the optimal settings which minimize incurred costs.

\section*{Results}

\subsection*{Criticality as the state of optimal control}

\begin{figure}[ht]
\centering
\includegraphics[width=\linewidth]{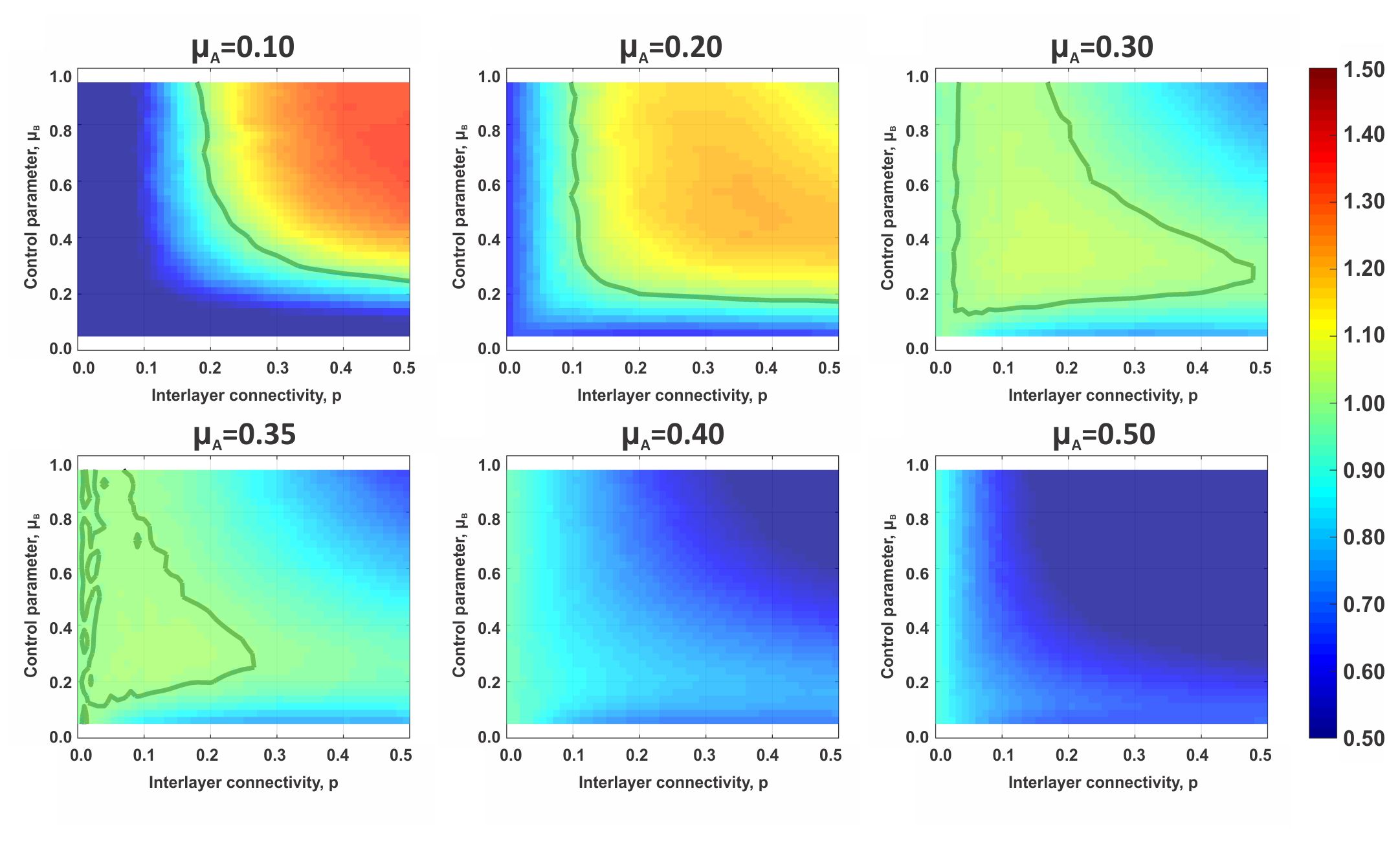}
\caption{Average cost of cascades experienced by layer A depends on control settings chosen by layer A and B, as well as on strength of interlayer coupling $p$. In all panels, the color denotes value of cost accrued by layer A normalized by the value obtained in fully uncontrolled case. Thick line denotes the value of normalized cost equal to $1$. In the case of $\mu_A=0.40$ and $\mu_A=0.50$ all values of normalized cost are smaller then 1, and thus the thick line is absent from those two panels. Size of an individual layer is $N=5000$ and all other parameters are as those used in Fig.\ref{fig:single_net}.}
\label{fig:colormaps}
\end{figure}

A wide range of complex systems, including biological, physiological, financial, ecological and social systems have been identified as operating at or near criticality \cite{thurner}. This dynamic state poised between order and disorder is believed to arise through evolutionary-like mechanisms as it is characterized by optimal trade-off between robustness and flexibility \cite{Chialvo2010}, the highest level of computational capabilities \cite{Hesse2014} and optimal dynamic range and memory \cite{Kinouchi2006}. Motivated by those observations we adopt a cost function $C(s)$ such that the average total cost of cascades is maximal for $\mu=\mu^\ast$, the value of the control parameter that corresponds to the critical state of the sandpile process. As illustrated on Fig.\ref{fig:single_net}b, a cost function $C(s)=\frac{1}{2}s^{3/4}$ results in a concave loss function. Since the gain decreases as a linear function of $\mu$, the average cost function $\langle C(\mu) \rangle$ retains the concave shape with a peak value near $\mu^\ast$. The values of the control parameter $\mu<\mu^\ast$ and $\mu>\mu^\ast$ result in a larger cost than one observed for $\mu^\ast$, denoting negative outcomes connected with the departure from the state of criticality. In the former case cascades are less frequent than in the classical sandpile model, however events encapsulating almost the entire system are more probable, resulting in an increase of the average cost. The latter range of parameters corresponds to a situation where increase in costs results from frequent small cascading events, which despite being all limited in size, on average result in bigger losses experienced by the system.  

The choice that particular control settings have on sandpile dynamics in a system of interconnected networks is shown in Figure \ref{fig:colormaps}. Consecutive panels correspond to increasing values of control parameter selected by layer A, $\mu_A$, while \textit{x-} and \textit{y-}axis of the individual color maps correspond, respectively, to a range of interlayer coupling $p$ and control parameter of layer B, $\mu_B$. Values denoted by the color maps represent the average normalized cost experienced by the layer A, $\langle C_{norm}^A(\mu_A,\mu_B,p) \rangle$, to illustrate how control settings selected by that layer are influenced by control chosen in layer B, as well as how the interlayer connectivity is affecting those parameters. 
The normalization is defined with respect to the native sandpile model
\begin{equation}
   \langle C_{norm}^A(\mu_A,\mu_B,p) \rangle=\frac{\langle C^A(\mu_A,\mu_B,p) \rangle}{\langle C^A(\mu^\ast,\mu^\ast,p)\rangle}
   \label{Cnorm}
\end{equation}
The thick line on the panels corresponds to $\langle C^A(\mu_A,\mu_B,p) \rangle=\langle C^A(\mu^\ast,\mu^\ast,p) \rangle$ and values $\langle C_{norm}^A(\mu_A,\mu_B,p)\rangle >1$ denote cases where the control strategy selected by layer A, combined with the control set by layer B, results in a better outcome than when $\mu_A=\mu_B=\mu^\ast$. Similarly, the values $\langle C_{norm}^A(\mu_A,\mu_B,p)\rangle <1$ denote choice of control that is more detrimental to the system than uncontrolled sandpile configuration. 

\begin{figure}[!ht]
\centering
\includegraphics[width=\linewidth]{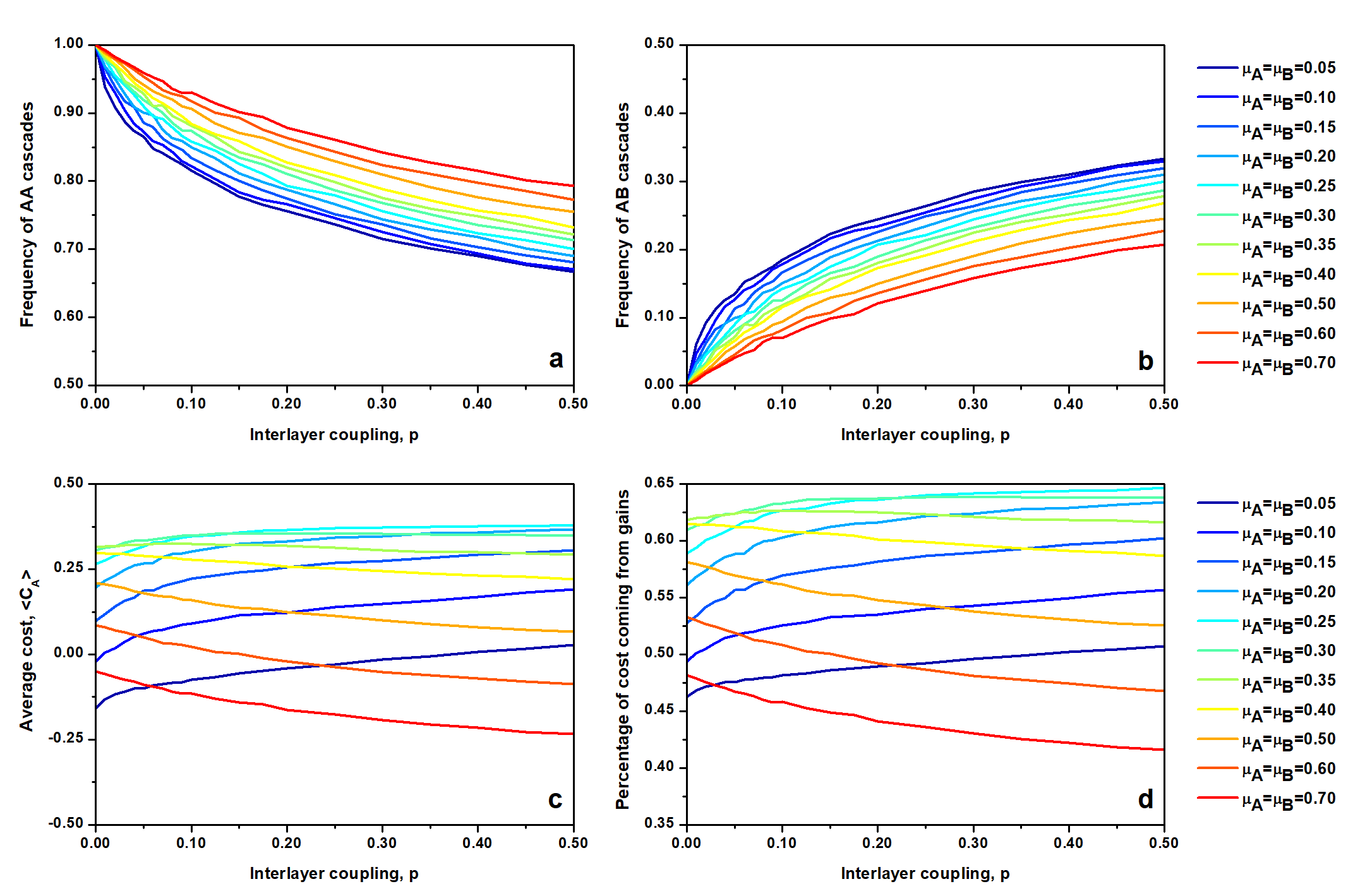}
\caption{ The case of matching control settings, $\mu_A=\mu_B$, provides insight into the behavior of controlled sandpile dynamics showed in Fig.\ref{fig:colormaps}. (a-b) Interlayer coupling modifies the frequency of observed cascades, with AB cascades (cascades that start in layer A and spill over into layer B) becoming more frequent as coupling $p$ increases. (c) Average cost of failures occurring in layer A is a function of interlayer coupling as well. However the functional dependence varies with $\mu$, and for $\mu<\mu^\ast$ cost increases with $p$, while for $\mu>\mu^\ast$ the relation is opposite. (d) The balance between gains caused by the avoidance of cascades and losses brought by local or inflicted cascades follows the behavior of the average cost. All parameters are as those listed in Fig.\ref{fig:colormaps}. }
\label{fig:CA_CB_same}
\end{figure}

Figure \ref{fig:colormaps} shows that the strategy chosen by layer A can result in three broadly defined outcomes. First, the condition of $\mu_A<\mu^\ast$ is the only one that leads to higher benefits than the classical sandpile case. Matched with layer B operating in the regime of $\mu_B>\mu^\ast$ and increasing interlayer coupling $p$, the layer A experiences positive effects of working in an interconnected system. We observe an opposite behavior for $\mu_A>\mu^\ast$, when any configuration results in outcomes worse than ones obtained for $\mu_A=\mu^\ast$. In particular the presence of layer B has a negative effect on the sandpile dynamics observed in layer A. The negative effect increases with the value of $\mu_B$ and interlayer coupling $p$. Finally the control in layer A set close to the native sandpile dynamics, $\mu_A \approx \mu^\ast$, requires either low interlayer coupling matched with wide range of settings chosen by layer B or stronger coupling $p$ followed by $\mu_B \approx \mu^\ast$, in order to see positive outcomes. 

In order to explain above observations we focus on a case where both layers operate with the same control setting, $\mu_A=\mu_B$. Such simplification of the dynamics allows us to investigate in detail how the average cost acquired by layer A is affected by the presence of a second layer and how varying control settings affect gains and losses incurred in the system. In order to determine the effects of interlayer coupling, we distinguish two types of cascading events: local ones, which are contained to the layer in which a cascade has started (denoted as AA or BB events) and inflicted ones, when a cascade spreads to the other layer (referred to as AB or BA events). Figures \ref{fig:CA_CB_same}a,b illustrate how frequency of both types of events varies with parameters of the system. With increasing interlayer coupling $p$ the local cascades become less frequent, as connections between layers facilitate spread of failures. However the strength of this effect depends on the control parameter. In systems with low values of $\mu$ local cascades are less probable, since the sandpile dynamics operates in a supercritical regime leading to cascading events encapsulating an entire layer. Thus it is more likely that a massive cascade, progressing through the layer in which it originated, will get transmitted to the opposite network rather than stay contained locally. Additionally, in the $\mu_A=\mu_B$ case considered here, the second layer is characterized by a supercritical dynamics as well. Since this state corresponds to a near saturation of the network with sand, where a significant number of nodes are near their capacity, a cascade originating from the first layer has optimal conditions for spreading into the second layer. The opposite is true for systems operating with high values of $\mu$, where sandpile dynamics is reminiscent of a subcritical state. Here both layers operate in a regime where large cascades are less likely, effectively limiting the chance for spread of failures across layers. 

The frequency of AA and AB events is one of the variables defining the average cost experienced by a network (see Eq. \ref{average_cost}). In particular the chance of AB cascades corresponds to the rate $\theta$, denoting the effect of interlayer coupling on the local sandpile dynamics. The average cost incurred by layer A (see Fig. \ref{fig:CA_CB_same}c) retains the concave shape for any $p$, being low for low values of $\mu$, then increasing for intermediate values of control and finally decreasing for high control settings. Furthermore, the value of $\mu=\mu^\ast$ is a transition point between two regimes, one in which average cost increases with interlayer coupling ($\mu<\mu^\ast$) and second one where higher interlayer coupling leads to reduction in cost ($\mu>\mu^\ast$). This change is derived from the shape of the cost function which is not symmetric with respect to $\mu^\ast$. As shown on Fig. \ref{fig:single_net}b, the difference between the loss and the total cost decreases as a function of $\mu$ and the total cost incurred at larger $\mu$ is mostly due to losses rather than gains. At low values of $\mu$, the selected control strategy and gains associated with it are responsible for higher percentage of total cost. 

\begin{figure}[!ht]
\centering
\includegraphics[width=\linewidth]{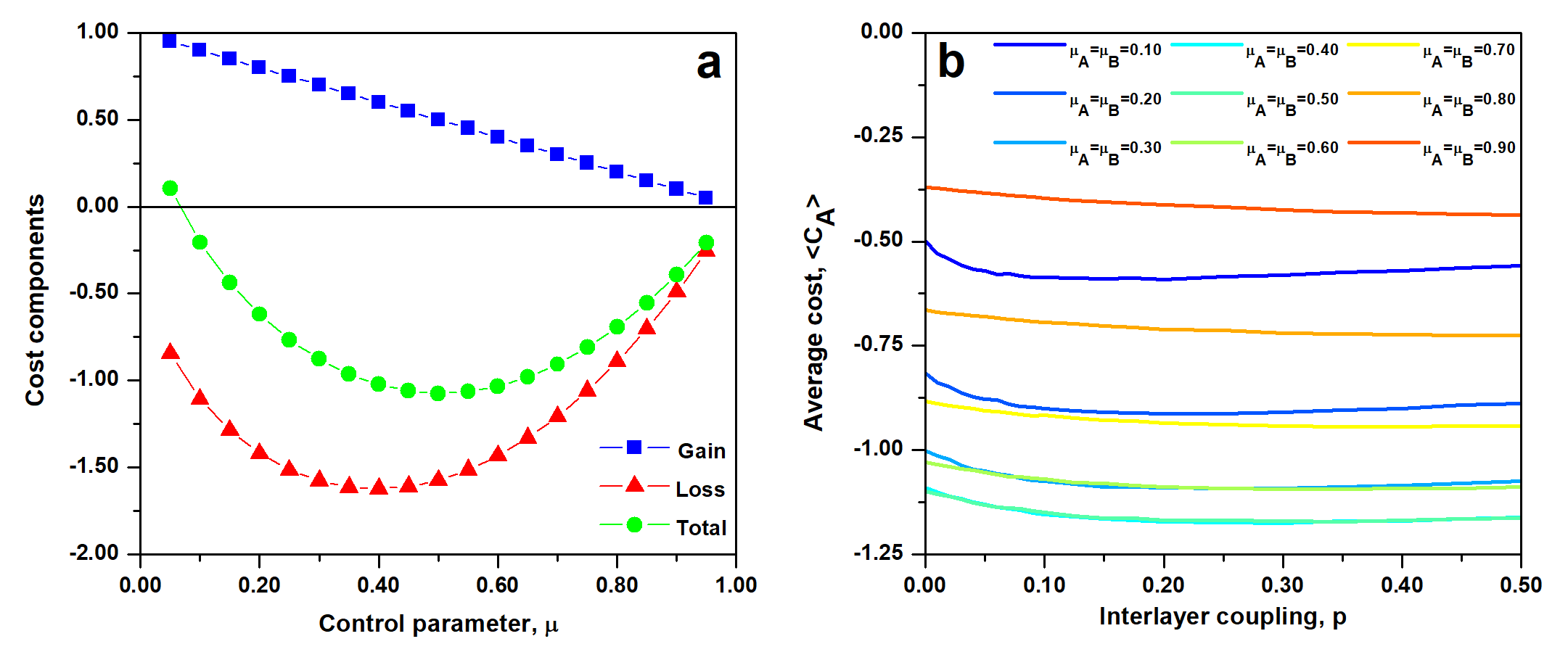}
\caption{ A cost function $C(s)\sim s^{3/4}(1-\mu^2)$ captures the idea that both strategies: one of avoiding cascades ($\mu \ll \mu^\ast$) and one of allowing only small cascading events ($\mu \gg \mu^\ast$), should result in small average cost, as rare large events as well as frequent small events might be associated with similar cost or cost versus risk trade-off. Panel (b) demonstrates how this modified cost definition affects the averaged total cost experienced by layer A in the case of $\mu_A=\mu_B$ control settings. The values presented here are not normalized as in Fig. \ref{fig:CA_CB_same}. They reflect absolute value of average cost $\langle C(\mu,\mu,p) \rangle$.}
\label{fig:Diff_cost}
\end{figure}

Thus in the regime $\mu<\mu^\ast$, the total cost increases with $p$, since coupling to a second layer allows to reduce load, which leads to reduction in the chance of largest cascades normally observed in A, in turn reducing losses. High frequency of cascades present in the regime of $\mu>\mu^\ast$ causes an opposite effect, since here the system is penalized by cascades of any size, and an increasing connectivity drives the probability of inflicted cascades.The gain associated with selected control value remains relatively constant as a function of $p$, as demonstrated by Fig. \ref{fig:CA_CB_same}d. Since the fraction of the average cost that is associated with gains follows the same dependence as the average cost, its absolute value must be independent of $p$, especially for low values of $\mu$.

Therefore the high values of normalized average cost seen for $\mu_A<\mu^\ast$ at high values of $\mu_B$ and $p$ (Fig. \ref{fig:colormaps}) stem from the fact that this particular configuration allows supercritical layer A to take advantage of  the subcritical layer B by depositing excess load and thus reducing the size of cascades generated in A. At the same time system B produces mostly events of small size, which do not often propagate through interlayer connections affecting layer A. From this perspective the layer A plays a role of a parasite taking advantage of the control setting adopted by layer B, while layer B behaves like a host organism, experiencing the interlayer coupling negatively. At high values of $\mu_A$ the optimal configuration for layer A is that of low connectivity and/or one of layer B operating in supercritical regime. Both cases limit the probability of inflicted cascades, either structurally through restricted number of interlayer connections, or dynamically by coupling to a system in which cascades are rare. 

\subsection*{Maximizing cost at extreme control settings}

\begin{figure}[ht]
\centering
\includegraphics[width=\linewidth]{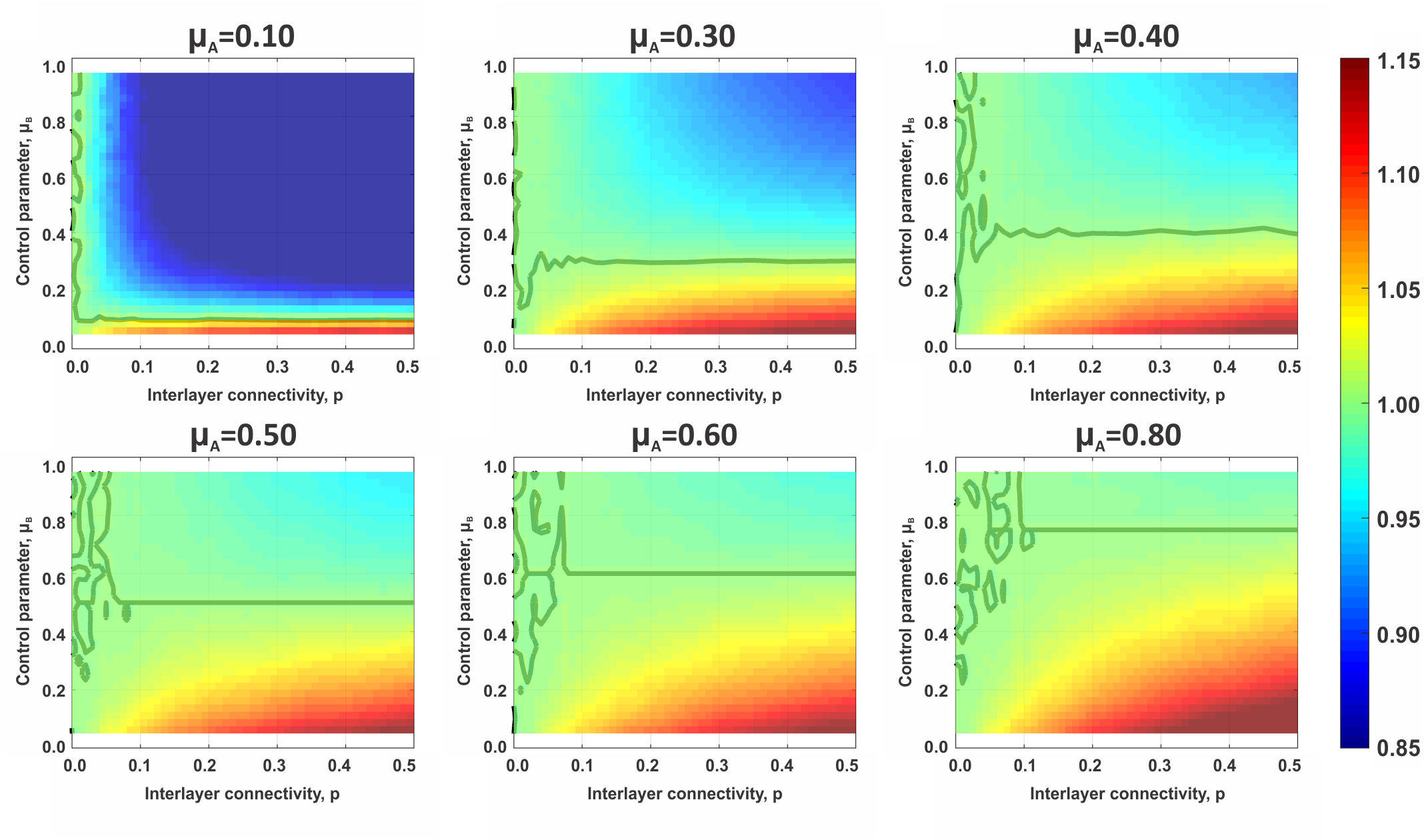}
\caption{ Greedy control of cascading failures needs to take into account control settings chosen by the opposite layer as well as strength of the interlayer coupling $p$. For any $\mu_A$ only $\mu_B<\mu_A$ settings chosen by layer B lead to positive outcomes for layer A. In all panels, the color denotes value of cost accrued by layer A normalized by the value obtained in the $\mu_A=\mu_B$ case (see Eq. \ref{Cnorm2}). Thick line denotes the value of normalized cost equal to $1$.}
\label{fig:Diff_cost_maps}
\end{figure}

In the previous section we discussed a case where the optimal control configuration, one associated with maximal benefits, corresponds to the critical dynamics of the sandpile model. That selection was motivated by the fact that numerous complex systems demonstrate the behavior reminiscent of a critical state, suggesting it being advantageous over other dynamical settings. Here however we set to analyze the dynamics from the perspective of a controller, one ultimately responsible for the damages caused by cascading failures propagating through the system. In this new framework, it is natural to expect that the controller prefers configurations limiting occurrences of cascades all together or ones that lead to cascades of small size. Thus we define a new cost function, one which reflects those preferences. The cost function of the form $C(s)\sim s^{3/4}(1-\mu^2)$ has a convex shape with a minimum near $\mu^\ast$ and two arms increasing as $\mu \rightarrow 0$ and $\mu \rightarrow 1$ (see Fig. \ref{fig:Diff_cost}a). Such a choice reflects the expectation of a controller that the regulatory setting leading to few cascades ($\mu \ll \mu^\ast$) or one leading to small cascades ($\mu \gg  \mu^\ast$) should be beneficial to the system. 

The convex shape of the cost function $C(s)$ is retained in the average cost $\langle C(s) \rangle$ incurred by a layer of the two-network system (see Fig. \ref{fig:Diff_cost}c), with only weak dependence on interlayer coupling $p$. In the case shown on the Fig. \ref{fig:Diff_cost}c, $\mu_A=\mu_B$, the extreme values of $\mu$ incur almost $50 \%$ smaller cost than that experienced at intermediate values of $\mu$. Next we relax the condition of $\mu_A=\mu_B$ to the whole range of possible control values and we demonstrate that for any value of $\mu_A$ the optimal configuration is coupling with layer B operated in $\mu_B < \mu_A$ regime. As shown on Fig. \ref{fig:Diff_cost_maps} this positive effect increases with interlayer coupling $p$. Here we need to note that the values shown on the color maps represent different normalization than one defined in Eq. \ref{Cnorm}. In order to address the issue of a control mechanism designed globally for a system of interconnected networks, we use the case $\mu_A=\mu_B$ as reference, and thus values shown in Fig.\ref{fig:Diff_cost_maps} represent 
\begin{equation}
   \langle C_{norm}^A(\mu_A,\mu_B,p) \rangle=\frac{\langle C^A(\mu_A,\mu_B,p) \rangle}{\langle C^A(\mu_A,\mu_A,p)\rangle}.
   \label{Cnorm2}
\end{equation}
Observed effects directly relate to the fact that dense interlayer connections facilitate spread of failures across layers of the system, and depending on the dynamical states of individual layers, the effects of the coupling can be either positive or negative. Layer operating in supercritical regime, $\mu_A \ll \mu^\ast$, generates few cascades and the coupling to a similarly supercritical network allows it to dissipate excessive load, thus decreasing the components of cost originating from losses. On the contrary coupling to a subcritical system leads to an increase in losses due to cascades occurring more easily in that layer and being transmitted through interlayer links. 

\section*{Discussion}

The dynamics of natural and man-made networks is usually nonlinear, making those systems complex not only with respect to their structure, but also with respect to their functional characteristics. Nonlinearieties, especially ones arising from feedback interactions, make it difficult to predict the impact of external perturbations on the system, and thus present severe obstacles to designing successful control protocols. In particular, structural cross-layer connections between networks can lead to unique phenomena not observed in the individual systems \cite{Havlin2010}. In this paper we set out to approach the problem of controlling dynamics on interconnected networks, focusing on regulating the dynamics of cascading failures. By formulating the control problem in terms of probability of occurrence of a cascade, we extent the dynamics of the sandpile model into both subcritical and supercritical regions. This allows us to test a variety of control strategies, from avoiding cascading events, through classical critical sandpile dynamics to limiting the size of failures. 

We demonstrate that coupling between networks introduces dependencies between their dynamical states, leading to coupling between control settings adopted by controllers operating in respective layers. Thus actions selected by controllers having in mind maximizing benefits and reducing losses in the system they operate become affected by each other. The interlayer coupling causes failures to propagate from one system to the other, leading to positive or negative side effects, potentially turning the strategy optimal for an individual layer into a suboptimal configuration. In this paper we demonstrate that the most generic control strategy is to position one's system in a diametrically different dynamical state with respect to it's partner. Additionally this effect is mediated by the density of interlayer connections, where in some dynamical configurations high connectivity is beneficial, while in others detrimental to interacting systems. Altogether we illustrate the impact of greedy, locally-designed control on the system's performance, demonstrating that more global, cooperative control protocols are needed in networked settings. 

\section*{Methods}
The sandpile dynamics outlined above is evaluated on random regular networks with degree $k=4$. Each network has $N=5000$ nodes and the dissipation parameter of the sandpile model is set to $f=0.05$. The probability distribution of cascade sizes observed in the individual layers is evaluated by iterating the sandpile dynamics for $2 \times 10^6$ time steps. At each time step a layer on which a grain of sand is being deposited is chosen at random and size of a cascade observed in each layer resulting from that deposition is recorded. The probability distribution of cascades observed in layer A (B), $P(s_A) (P(s_B))$, is estimated through logarithmically binned histogram. This probability distribution is then used to calculate the average cost according to Eq. \ref{average_cost}.  


\section*{Author contributions statement}
M.T. designed the study and conducted numerical simulations. All authors wrote the manuscript. 

\section*{Additional information}
\textbf{Competing interests} The authors declare no competing interests. 

\end{document}